# Experimental Study of Boron-coated Straws with a Neutron Source

Zhaoyang Xie[1,2,3], Jianrong Zhou[2,3,5;*], Yushou Song[1;1)], Liang Sun[4;2)], Zhijia Sun[2,3], Bitao Hu[5], Yuanbo Chen[2,3]

[1] Key Discipline Laboratory of Nuclear Safety and Simulation Technology, Harbin Engineering University, Harbin 150001, China
[2] Institute of High Energy Physics, Chinese Academy of Sciences (CAS), Beijing 100049, China
[3] Dongguan Neutron Science Center, Dongguan 523803, China
[4] Proportional Technologies, Inc., 8022El Rio Street, Houston, TX, 77054, USA
[5] School of Nuclear Science and Technology, Lanzhou University, Lanzhou 730000, China

**Abstract:** Multiple types of high quality neutron detectors are proposed for the first phase of the China Spallation Neutron Source (CSNS), which will be commissioned in 2018. Considering the shortage of $^3$He supply, a detector module composed of 49 boron-coated straws (BCS) was developed by Proportional Technologies Inc. (PTI). Each straw has a length of 1000 mm and a diameter of 7.5 mm. Seven straws are tightly packed in a tube, and seven tubes are organized in a row to form a detector module. The charge division method is used for longitudinal positioning. A specific readout system was utilized to output the signal and simultaneously encode each straw. The performance of this detector module was studied using a moderated $^{252}$Cf source at the Institute of High Energy Physics (IHEP). The signal amplitude spectrum indicates its n-gamma discrimination capability. Despite the complex readout method, a longitudinal resolution of 6.1±0.5 mm was obtained. The three-dimensional positioning ability qualifies this BCS detector module as a promising neutron scattering spectrometer detector.

## 1. Introduction

Neutron scattering [1] is an effective method to observe the microstructures of materials and biological molecules. The China Spallation Neutron Source (CSNS) being constructed in Dongguan will provide high quality neutron beams for such applications. In addition to the neutron source, a neutron scattering spectrometer composed of large area of position-sensitive neutron detectors is another essential instrument. The existing spectrometers use different kinds of detectors, including multi-wire proportional chambers ($^3$He) [2], scintillators [3] and $^3$He linear position sensitive detectors (LPSD) [4]. Among these detectors, $^3$He-based LPSDs were used most widely due to their outstanding performance. For example, the original MWPCs in SANS2d were replaced by $^3$He LPSDs in 2014. As a result, efficiency improved by 30-40%, and a higher counting rate capability was achieved [5].

On the other hand, the issue of increased demand and limited supply of $^3$He [6] has caused a serious shortage of this rare gas. Multiple $^3$He-replacement neutron detection technologies have been established in recent years. The boron coated straw (BCS) technology is a promising candidate that can be manufactured easily and economically for large area applications. It shows many advantages including high count rate capability and n-gamma discrimination ability. Furthermore, the BCS is a gas detector working under proportional mode just like a $^3$He-based LPSD, and it shares many features in common with the LPSD. The existing maintenance techniques and readout system for the LPSD can be conveniently transferred to the BCS.

For neutron imaging purposes used in homeland security, BCSs with a diameter of 4 mm were developed and studied by Proportional Technologies Inc. (PTI) [7-14]. In this report, straws with a diameter of 7.5 mm were developed by PTI to maintain geometric compatibility with the $^3$He LPSDs used in existing neutron scattering spectrometers such as SANS2d [5] and BIO-SANS [15]. Seven straws were sealed in a tube, and seven tubes were packed in a row to form an expandable detector module. The performance of this module was studied by a moderated $^{252}$Cf neutron source at the Institute of High Energy Physics (IHEP). To improve the signal processing quality and reduce the number of readout channels, the multiplex readout method was used [16].

In section 2, the experimental setup is described in detail. In section 3, the n-gamma discriminating and positioning abilities are discussed along with the experimental results. Lastly, a summary is given in section 4.

## 2. Experimental setup

A BCS is 1000 mm long (effective 800 mm) and 7.5 mm in diameter. The wall of the cathode tube is coated with one-micron thick 96% enriched boron carbide ($B_4C$). A Stablohm wire with a diameter of 20 microns is tensioned in the center of each straw as an anode. A bias voltage of 1051 V is applied between anode wire and cathode tube, which makes the straw work in proportional mode. A mixture of argon and carbon dioxide with a ratio of (90%/10%) is used as the working gas (0.7 atm). Seven BCSs are hexagonally configured in a tube, and seven BCS tubes are grouped to form a detecting module. The seven tubes are labeled tube #1 to tube #7, and the

---
* Joint first author
1) Corresponding author: songyushou80@163.com
2) Corresponding author: lsun@proportionaltech.com

straws within a tube are labeled straw #1 to straw #6, as shown in Fig. 1

The longitudinal (along the anode wire direction) position is determined by the charge division method. The two ends of each straw are labeled as A and B. To reduce electronic routes and identify every straw, a particular readout design is used, as shown in Fig. 1. Each straw has an independent preamplifier. In the A end of the setup, the signals from straws with the same index across seven tubes go to the same main amplifier and the same ADC. To make the diagram more clear, only the readout of straws labeled as 6 are plotted. In the other end, the signals coming from seven straws in a tube go to the same main amplifier, then to the same ADC. In Fig. 1, only the readout for "Tube 7" is plotted.

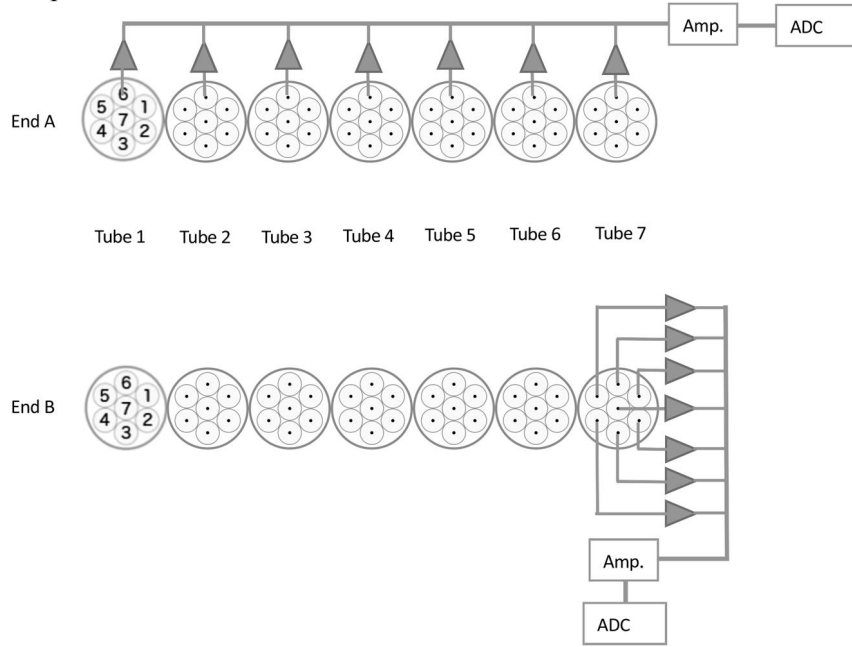

**Figure 1. The schematics of the readout system for this detector module, where the tubes are labeled tube #1 to tube #7, and the straws within a tube are labeled straw #1 to straw #7.**

A $6.3 \times 10^5$ Bq $^{252}$Cf was used as the neutron source in this experiment. The source itself is surrounded by lead sphere shielding to suppress the gamma rays. The neutron moderator layer consists of 35 cm thick paraffin. Borax of 11 cm thick is placed at the outermost layer for dumping of the moderated neutrons. A neutron beam collimator tunnel with a diameter of 10 cm is prepared, starting from the lead sphere and continuing through the neutron moderating and dumping materials. The BCS detector module was positioned immediately next to the collimator tunnel. Because most neutrons coming through the tunnel are fast, a cylindrical moderator with a thickness of 10 cm is placed in the tunnel to thermalize the neutrons in this experiment. To obtain the longitudinal positioning resolution, a collimator made of cadmium with a 1 mm wide slit is attached behind the cylindrical moderator.

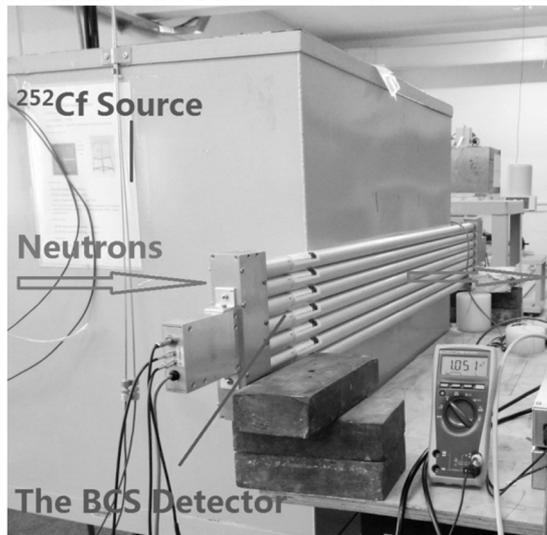

**Figure 2. The picture of the experimental setup at IHEP.**

## 3. Results and discussion
### 3.1 Pulse Amplitude Spectrum and Gamma Rejection

The pulse amplitude spectrum of a straw is shown in Fig. 3(a). Although most of the gamma rays from the $^{252}$Cf source are eliminated by the shields, there always remain some escaping gamma rays captured by the straws. There are also some background gamma rays that enter the spectrum. A gamma peak is hence formed on the left of the spectrum.

To understand the neutron spectrum comprehensively, the following analysis is provided. An incident neutron is captured by the B$_4$C coated layer through the following reactions [11, 17].

$$^{10}_{5}B + n \rightarrow ^{7}_{3}Li + \alpha + Q_1 \quad (Q_1 = 2.792 MeV, 94\%)$$
$$^{10}_{5}B + n \rightarrow ^{7}_{3}Li^* + \alpha + Q_2 \quad (Q_2 = 2.310 MeV, 6\%)$$
(1)

The reaction products $^7$Li and alpha fly back to back as the incident energy is too small in contrast to the reaction energy $Q$. Due to the geometry of the cathode, only the produced particle flying inward has a chance to escape the coated B$_4$C film and ionize the working gas molecules. Different neutron absorption depths in the film and different flying directions make the residual kinetic energy of an escaping particle vary continuously, hence the continuous energy deposition. If the neutron absorption position is close to the surface of the coated film and the flying direction is proper, the produced particle may have enough energy to penetrate the working gas. In an alternate scenario, a particle may not fully deposit its kinetic energy. The straws work in proportional mode under approximately 1050 V, and therefore continuous energy deposition produces a continuous signal amplitude spectrum. To fully understand the spectrum, a simulation based on Geant4 [18] was performed. As shown in Fig. 3 (b) we obtained the simulated spectrums contributed from $^7$Li, alpha and the total without considering the energy resolution. It is consistent with the experimental result. In further analysis, a threshold of 300 ch (around the gamma tail of 30 keV) used to eliminate the gamma background was set as indicated in Fig. 3 (a).

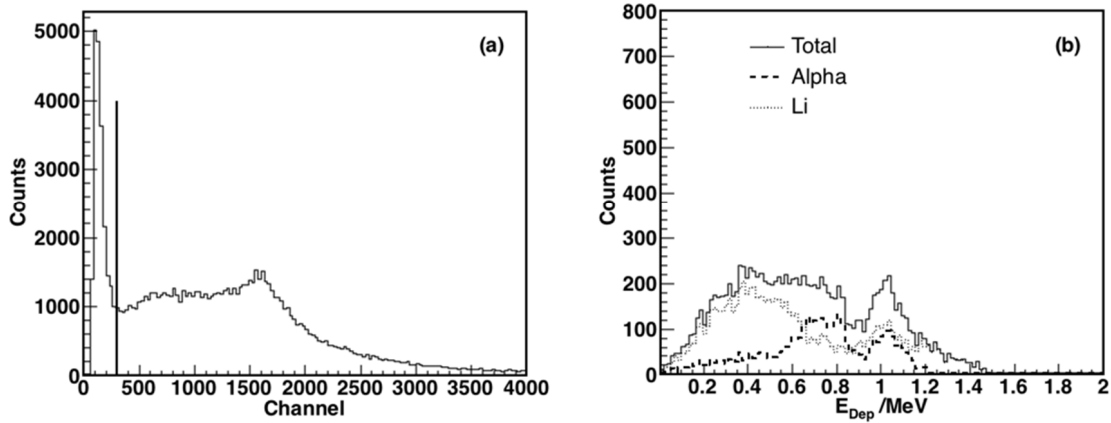

**Figure 3. Pulse amplitude spectrum of a straw in this experiment (a) and the simulated energy deposition of neutron-produced alpha and 7Li (b).**

### 3.2 Longitudinal resolution

To study the longitudinal resolution of a straw, an additional cadmium plate with a slit of 1 mm functioning as a collimator was installed. The charge division method was used to determine the neutron absorption positions along a straw. As an example, the neutron-absorption position spectrum of straw #2 in tube 4 is given in Fig. 4 by the formula

$$x = \frac{Q_B}{Q_A + Q_B}, \quad (2)$$

where $Q_A$ and $Q_B$ are the charge quantities of end A and end B of the straw, respectively.

The neutrons detected by the detector module can be classified into three groups: i) neutrons from the background and leaking from the source (i.e., penetrating the moderating and dumping layers); ii) neutrons traveling along the beam tunnel and penetrating the cylinder moderator fixed in the tunnel and cadmium collimator plate; and iii) neutrons passing through the slit of the cadmium collimator. The neutron flux of type i) is random along the straw axis and their directions are isotropic. Therefore, they form a uniform background in the longitudinal positioning spectrum. Although the moderator cylinder is installed in the beam tunnel, not all the neutrons are sufficiently thermalized and cut off by the cadmium collimator. These neutrons penetrating the cadmium collimator are of type ii). These neutrons form a Gaussian-like structure upon the uniform background. Its width is a little bit broader than the beam tunnel diameter due to the wall effect of the tunnel and the momentum direction straggling of the neutrons.

The peak in the spectrum is due to the neutrons of type iii) and sits on the background composed of the two structures described above, We extracted this peak, which interprets the approximate position resolution, by subtracting a fitted uniform and a Gaussian distribution. With a Gaussian distribution fitted on this peak, the FWHM=6.1±0.5 mm was obtained. This longitudinal resolution is close to the result of the straw with a diameter of 4 mm [7, 9] and is much better than that from the earlier testing results of straws with the same diameter [14].

The longitudinal resolution is constrained physically by the ranges of neutron-produced alpha and $^7$Li and the avalanche scale in the working gas. The present resolution nearly reaches the ideal resolution and mostly benefits from the newly designed readout method. Compared with the conventional delay-line readout method [12], signal delaying and summing of the dual-end signals of a delay line are thrown out, which may induce uncertainty.

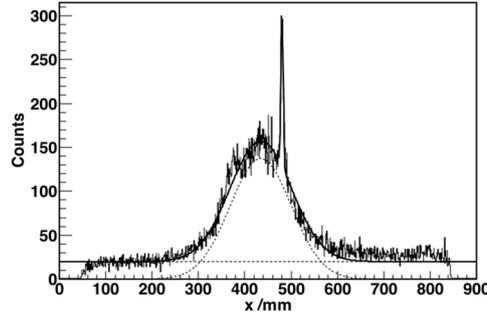

Figure 4. The longitudinal position spectrum (histogram). The dotted lines are the fitted two kinds of background, and the solid line is the full fitting of the spectrum.

### 3.3 Three-dimensional positioning

The neutron detection efficiency of a gas detector with a neutron converter is usually not high due to the limitation of the thickness of the converting film. However, certain efficiency is demanded by a neutron scattering spectrometer. Packing detector units to form a detector array is an effective way to improve the efficiency. Three-dimensional positioning of the neutron absorption promises the neutron scattering angle is reconstructed correctly.

In addition to the longitudinal positioning being determined by charge division, the positioning of the two dimensions perpendicular to the straw wires is determined by decoding the straw indexes from readout system. We obtained the three-dimensional neutron absorption positions from the experiment (without cadmium collimator) as shown in Fig. 5(a). In this graph the x-axis is along the straw wire direction and the z-axis orients to the beam direction. The origin is set on the geometric center of the detector module (middle point of the anode wire of straw #7 in tube #4). The diameter of a straw is 7.5 mm, and the uncertainty of positioning in the y and z directions is 7.5 mm, which is similar to that of the longitudinal direction. A straw may be fabricated longer to make the detector module area larger. Therefore, this detector module composed of straws is capable of expanding to fit a large spectrometer with sufficient detection efficiency. Projecting the three-dimensional positioning histogram to the xoy plane, we get the beam profile as shown in Fig. 5 (b). This indicates that this detector module also has the potential to be a neutron imaging detector.

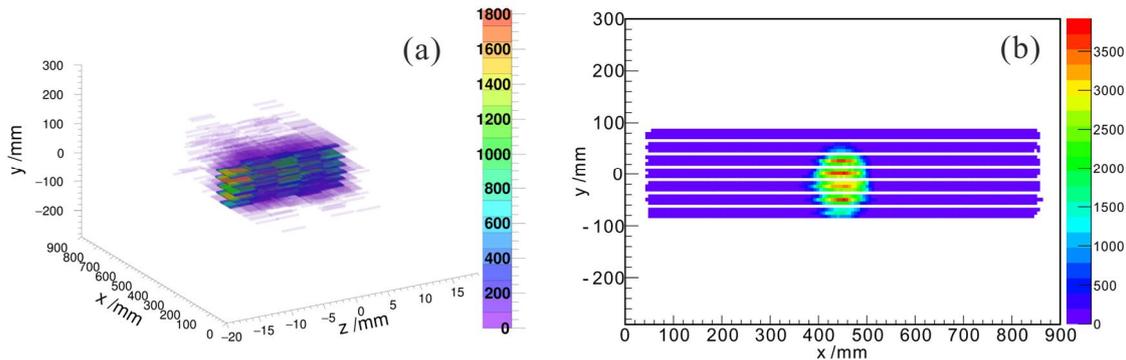

Figure 5. The three-dimensional and two-dimensional positioning performance of neutron absorption by the detector module.

### 4. Summary

To find a proper $^3$He free neutron detector for the neutron scattering spectrometers of CSNS, a particular BCS was developed in PTI. For convenience in comparing its performance with those of $^3$He-based LPSDs used in many neutron scattering spectrometers, we employed similar dimensions for the straws as the LPSDs. Based on these straws, a neutron detection module with independent functions was proposed. An experimental study on the performance of this neutron detection module was carried out at IHEP. In the signal amplitude spectrum, the events of neutrons are distinguishable from that of gammas, which guarantees the ability of the detector to eliminate the entering gammas. The best longitudinal resolution achieved was 6.1±0.5 mm. The transverse positioning resolution is determined by the diameter of a straw, which is similar to the longitudinal resolution. In this way, the three-dimensional positioning of neutron absorption is realized by this neutron detector module. In a neutron scattering spectrometer, the detector module can be extended to a scale which satisfies certain detecting efficiencies without corrupting the scattering angle accuracy. In this experiment, we discovered a series of features in this detecting module suitable for acting as a detector unit of a neutron scattering spectrometer. More comprehensive studies, such as aging and working stability, are required before future practical use.

**Acknowledgements**

This work is supported by National Natural Science Foundation of China [grant numbers 11205036, 11405191], CAS [grant numbers YZ201512] and the Fundamental Research Funds for the Central Universities of China [grant numbers HEUCF131501].